# On the Performance of MPI-OpenMP on a 12 nodes Multi-core Cluster

Abdelgadir Tageldin Abdelgadir[1], Al-Sakib Khan Pathan[1]*, Mohiuddin Ahmed[2]

[1] Department of Computer Science, International Islamic University Malaysia,
Gombak 53100, Kuala Lumpur, Malaysia

[2] Department of Computer Network, Jazan University, Saudi Arabia

abddu1@gmail.com , sakib@iium.edu.my , ahmed255555@yahoo.com

**Abstract.** With the increasing number of Quad-Core-based clusters and the introduction of compute nodes designed with large memory capacity shared by multiple cores, new problems related to scalability arise. In this paper, we analyze the overall performance of a cluster built with nodes having a dual Quad-Core Processor on each node. Some benchmark results are presented and some observations are mentioned when handling such processors on a benchmark test. A Quad-Core-based cluster's complexity arises from the fact that both local communication and network communications between the running processes need to be addressed. The potentials of an MPI-OpenMP approach are pinpointed because of its reduced communication overhead. At the end, we come to a conclusion that an MPI-OpenMP solution should be considered in such clusters since optimizing network communications between nodes is as important as optimizing local communications between processors in a multi-core cluster.

## 1  Introduction

The integration of two or more processors within a single chip is an advanced technology for tackling the disadvantages exposed by a single core when it comes to increasing the speed, as more heat is generated and more power is consumed by those single cores. The word core refers as well to a processor in this new context and can be used interchangeably. Some of the famous and common examples of these processors are the Intel Quad Core; which is the processor our research cluster is based on, and the AMD Opteron or Phenom Quad-core. This aggregation of classical cores into a single "Processor" has introduced the division of workload among the multiple processing cores as if the execution was to happen on a fast single processor, this also introduced the need of parallel and multi-threaded approaches in solving most kinds of problems. When Quad-cores processors are deployed in a cluster, 3 types of communication links must be considered: (i) between the two processors on the same chip, (ii) between the chips in a same node, and (iii) between different processors in different nodes. All these communications methods need to be considered on such cluster in order to deal with the associated challenges [1], [2], [3].

---

*This work was supported by IIUM research incentive funds. Abdelgadir Tageldin Abdelgadir also has been working with MIMOS Berhad research institute.

The rest of the paper is organized as follows: in Section 2, we briefly introduce MPI and OpenMP and discuss performance measurement with High Performance Linpack (HPL), Section 3 presents the architecture of our cluster, Section 4 describes the research methodologies used, Section 5 records our findings and future expectations and Section 6 concludes the paper.

## 2 Basic Terminologies and Background

### 2.1 MPI And OpenMP

The Message passing models provide a method of communication amongst sequential processes in a parallel environment. These processes execute on the different nodes in a cluster but interact by "passing messages", hence the name. There can be more than a single process thread in each processor. The Message Passing Interface (MPI) [8] approach simply focuses on the process communication happening across the network, while the OpenMP targets inter-process communications between processors. With this in mind, it will make more sense to employ OpenMP parallelization for inter-process communications within the node and MPI for message passing and network communication between nodes. It is also possible to use MPI for each core as a separate entity with its own address space; this will force us to deal with the cluster differently though. With this simple definitions of MPI and OpenMP, a question arises whether it will be advantageous to employ a hybrid mode where more than one OpenMP and MPI process with multiple threads on a node so that there is at least some explicit intra-node communications [2], [3].

### 2.2 Performance Measurement with HPL

High Performance Linpack (HPL) is a well-known benchmark suitable for parallel workloads that are core-limited and memory intensive. Linpack is a floating-point benchmark that solves a dense system of linear equations in parallel. The result of the test is a metric called *GigaFlops* that translates to billions of floating point operations per second. Linpack performs an operation called LU Factorization. This is a highly parallel process, utilizing the processor's cache up to the maximum limit possible, though the HPL benchmark itself may not be considered as a memory intensive benchmark. The processor operations it does perform are predominantly 64-bit floating-point vector operations and uses SSE instructions. This benchmark is used to determine the world's top-500 fastest computers. In this work, HPL is used to measures the performance of a single node and consequently, a cluster of nodes through a simulated replication of scientific and mathematical applications by solving a dense system of linear equations.

In the HPL benchmark, there are a number of metrics used to rate a system. One of these important measures is *Rmax*, measured in Gigaflops that represents the maximum performance achievable by a system. In addition, there is also *Rpeak*, which is the theoretical peak performance for a specific system [4]; this is obtained from:

$$[N_{proc} * Clock\ freq * FP/clock] \qquad (1)$$

Where $N_{proc}$ is the number of processors available, *FP/clock* is the floating-point operation per clock cycle, *Clock freq* is the frequency of a processor in MHz or GHz.

## 3   The Architecture of our Cluster

Our cluster consists of 12 *Compute Nodes* and a *Head Node* as depicted in Figure 1.

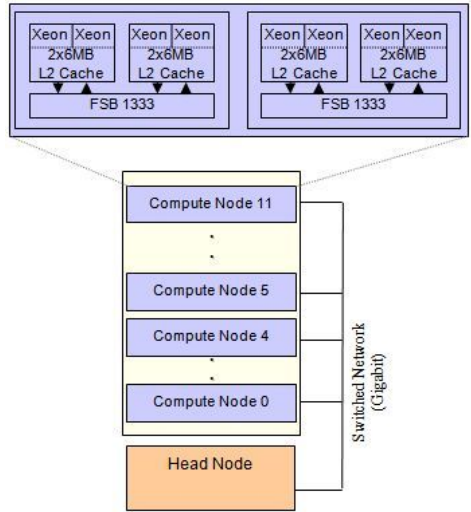

**Fig. 1.** Cluster physical architecture

### 3.1   Machine Specifications

The cluster consisted of two node types: a *Head Node* and *Compute Nodes*. The tests were run on the compute nodes only as the Head node was different in both capacity and speed. Its addition will increase the complexity of the tests. The Compute Node specifications shown in Table 1 are same. Each node has an Intel Xeon Dual Quad Core Processor running at 3.00 GHz. Note that the system had eight of the mentioned processors and that the sufficient size of cache reduces the latencies in accessing instructions and data; this generally improves performance for applications working on large amount of data sets. The Head Node specification (Table 2) was similar and had an Intel Quad Xeon Quad Core Processor running at 2.9 GHz.

**Table 1.** Compute Node processing specifications.

| Element | Features |
|---|---|
| processor | 0      (Upto 7) |
| cpu family | 6 |
| model name | Intel(R) Xeon(R) CPU    E5450  @ 3.00GHz |
| stepping | 6 |
| cpu MHz | 2992.508 |
| cache size | 6144 KB |
| cpu cores | 4 |
| fpu | yes |
| flags | fpu vme de pse tsc msr pae mce cx8 apic sep mtrr pge mca cmov pat pse36 clflush dts acpi mmx fxsr sse sse2 ss ht tm syscall nx lm constant_tsc pni monitor ds_cpl vmx est tm2 cx16 xtpr lahf_lm |
| bogomips | 6050.72 |
| clflush size | 64 |
| cache_alignment | 64 |
| address sizes | 38 bits physical, 48 bits virtual |
| RAM | 16GB |

### 3.2  Cluster Configuration

The cluster was built using Rocks 5.1 64-bits Cluster Suite, Rocks [9] is a Linux Distribution based on CentOS [10], it is intended for High Performance Computing systems. Intel 10.1 compiler suite was used; the Intel MPI implementation and the Intel Math Kernel Library were utilized as well. The cluster was connected to two networks, one used for MPI-based operations and the other for normal data transfer. As a side-note relevant to practitioners, a failed attempt was done with an HPCC version of the Linpack benchmark that utilized an OpenMPI library implementation; the results were unexpectedly low. Tests based on the OpenMPI configuration and subsequent planned test-runs were aborted.

**Table 2.** Head Node specifications.

| Element | Features |
|---|---|
| processor | 0 (*Upto 16*) |
| cpu family | 6 |
| model name | Intel(R) Xeon(R) CPU    X7350  @ 2.93GHz |
| stepping | 11 |
| cpu MHz | 2925.874 |
| cache size | 4096 KB |
| cpu cores | 4 |
| fpu | yes |
| flags | fpu vme de pse tsc msr pae mce cx8 apic sep mtrr pge mca cmov pat pse36 clflush dts acpi mmx fxsr sse sse2 ss ht tm syscall nx lm |

| | constant_tsc pni monitor ds_cpl vmx est tm2 cx16 xtpr lahf_lm |
|---|---|
| bogomips | 5855.95 |
| clflush size | 64 |
| cache_alignment | 64 |
| address sizes | 40 bits physical, 48 bits virtual |

# 4 Research Methodology

Tests were done in two main iterations, the first iteration was a single node performance measurement followed by an extended iteration that included all the 12 nodes. These tests consumed a lot of time; the cluster was not fully dedicated for pure research purposes as it was used as a production cluster as well, time was limited for test-runs. Our main research focused on examining to what extent, the cluster would scale, as it was the first Quad-core to be deployed at the site. In this paper, we focus on the much more successful test-run of the hybrid implementation of HPL by Intel for Xeon Processors. In each of the iterations, different configurations and set-ups were implemented; these included changing the grid topology used by HPL according to different settings. This was needed since the cluster contained both an internal grid – between processors – and an external grid composed of the nodes themselves. In each test trial, a configuration was set and performance was measured using HPL. An analysis of the factors affecting performance is recorded for each trial and graphs were generated to clarify process-distribution in the grid of processes.

## 4.1 Single Node Test

The test for a single node was done for all nodes. This is a precautionary measure to check whether all nodes are performing as expected since the cluster's performance in an HPL test-run is limited by the slowest of nodes. Table 3 shows the results from different nodes, the average is approximately 75.6 Gflops. This number can be calculated using Equation 2. In each node, there are Dual Xeon Quad Processors, making the theoretical peak performance equal to:

$$R_{peak} = 8*3*4 = 96 Gflops/node. \qquad (2)$$

But the maximum performance obtained was at an approximate average of 75.6 Gflops/node, this is the Rmax Value obtainable for a single node. The efficiency is calculated at 78.8%.

Table 3. Performance of Cluster Nodes

| **Node 1:** 7.517e+01 *Gflops*, **Node 2:** 7.559e+01 *Gflops*, **Node 3:** 7.560e+01 *Gflops*, |
| **Node 4:** 7.552e+01 *Gflops*, **Node 5:** 7.558e+01 *Gflops*, **Node 6:** 7.559e+01 *Gflops*, |
| **Node 7:** 7.557e+01 *Gflops*, **Node 8:** 7.560e+01 *Gflops*, **Node 9:** 7.537e+01 *Gflops*, |
| **Node 10:** 7.561e+01 *Gflops* **Node 11:** 7.557e+01 *Gflops* **Node 12:** 7.562e+01 *Gflops* |

Table 4 shows the parameters used for the single node test.

## 4.2 Multiple Nodes Test

The Multiple node test required many iterations to scale well and reach an optimal performance in the limited time the researcher had. The first thing put into consideration was the grid topology to be used in order to achieve good results. Several grids were proposed depending on the knowledge gathered from previous experiences; it is considered that in a cluster-wide test, attainment of high performance is dependent on the number of cores and the frequency of the processor being used on each node. Distribution of processes is crucial; a balanced distribution of processes will basically result in better performance.

**Table 4.** HPL configuration for Single Node test.

| Choice | Parameters |
|---|---|
| 6 | device out (6=stdout,7=stderr,file) |
| 1 | # of problems sizes (N) |
| 40000 | Ns |
| 1 | # of NBs |
| 192 | NBs |
| 0 | PMAP process mapping (0=Row-,1=Column-major) |
| 1 | # of process grids (P x Q) |
| 1 | Ps |
| 8 | Qs |
| 16.0 | threshold |
| 1 | # of panel fact |
| 0 1 2 | PFACTs (0=left, 1=Crout, 2=Right) |
| 1 | # of recursive stopping criterium |
| 4 2 | NBMINs (>= 1) |
| 1 | # of panels in recursion |
| 2 | NDIVs |
| 1 | # of recursive panel fact. |
| 1 0 2 | RFACTs (0=left, 1=Crout, 2=Right) |
| 1 | # of broadcast |
| 0 | BCASTs (0=1rg,1=1rM,2=2rg,3=2rM,4=Lng,5=LnM) |
| 1 | # of lookahead depth |
| 0 | DEPTHs (>=0) |
| 2 | SWAP (0=bin-exch,1=long,2=mix) |
| 256 | swapping threshold |
| 1 | L1 in (0=transposed,1=no-transposed) form |
| 1 | U  in (0=transposed,1=no-transposed) form |
| 0 | Equilibration (0=no,1=yes) |
| 8 | memory alignment in double (> 0) |

Generally, HPL is controlled by two main parameters that describe how processes are distributed across the cluster's nodes; these values P and Q are both critical benchmark-tuning parameters when producing good performance is required. P and Q should be as close to equal as possible, but when they are not equal; P should be less than Q. That is because when P is multiplied by Q, it actually gives the number of MPI processes to be used and how they are distributed across the nodes. In this cluster, there are several choices, such as *1x96, 2x48, 3x32, 4x24, 6x16, 8x12*. However, the network can affect performance, and in our case, the introduction of Multi-Cores within a single node; so different trials are needed to achieve best performance. Another parameter needed is N, which is the size of the problem to be fed to HPL. We have used the following formula as in [4] to estimate the problem size:

$$\sqrt{[((\sum M_{sizes})MB * 1000000000)/8]} \qquad (3)$$

This will give a value that will approximately be N, for example:

$$N = sqrt(12*16*1000000000) \sim= 154919$$

However, it is preferred not to take the whole result, we chose 140000 as N, giving more than 25% to other local system processes; this is to avoid the use of the virtual memory which will render the whole test-run useless. An overloaded system will use the swap area, and this will negatively affect the results of the benchmark. It is advisable to make full use of main memory, but at the same time avoid using the virtual memory.

The optimal performance was achieved with HPL input parameters as in Table 5:

**Table 5.** HPL configuration for 12 Nodes test.

| Parameter | Value |
|---|---|
| N | 140000 |
| NB | 192 |
| PMAP | Row-major process mapping |
| P | 6 |
| Q | 16 |
| RFACT | Crout |
| BCAST | 1ring |
| SWAP | Mix (threshold = 256) |
| L1 | no-transposed form |
| U | no-transposed form |
| EQUIL | no |
| ALIGN | 8 double precision words |

A first expectation was *3x32* or *4x24* will produce the optimal performance, but a *6x16* grid (Figure 2 and Figure 3) obtained the best performance at 662.2 Gflops, Performance increase is linear to some extent, but will not equal the overall absolute sum of 12 nodes that is 907 Gflops. This is acceptable, as a cluster's performance does not scale linearly in reality [1], thus the efficiency of the cluster is calculated at approximately 60%, which is satisfactory for a Gigabit-based cluster.

## 5 Observations, Discussions, and Future Expectations

By looking at the general topological structure of this cluster, we notice that different cores will be completing the same process in parallel, this leads to high network communication between the different nodes in such clusters. Moreover, processing speed tends to be faster than the Gigabit network's communication link speed available for the cluster. This will be translated into waiting time in which some cores may become idle. In preliminary test-runs, we opted to use an MPI-only approach based on our previous experiences with clusters, the results were disappointing, reaching a maximum of approximately 205Gflops. An option was proposed to run the Linpack benchmark test using Intel's MPI library in its hybrid mode, this version featured an MPI-OpenMP implementation of HPL, it uses MPI for the network communication while utilizing OpenMP for local communication between cores. This approach seemed more appropriate for a Multi-Node cluster and the results previously presented in this paper are based on a Hybrid implementation of HPL benchmark. The direct effect of this was a fully saturated network as well as fully utilized processors. [5], [6].

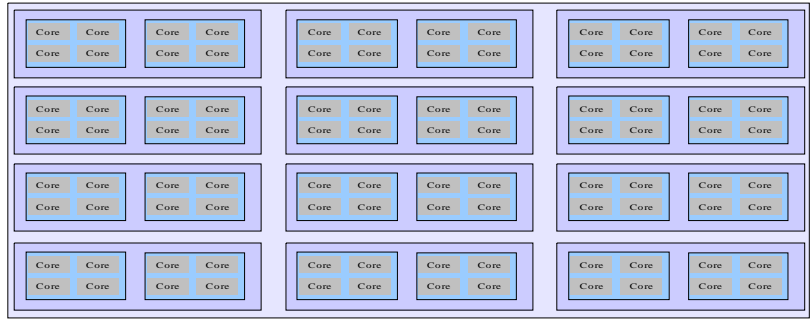

**Fig. 2.** Physical view of 6x16 cores grid

**Fig. 3.** Abstract view of MPI process distribution on the 6x16 grid

Another factor observed which affected the performance of this cluster is the network, the current set-up of the cluster includes two networks, one is used solely for MPI traffic, which is the network that obtained the highest possible result. From these findings, it is recommended that multi-core clusters deployed for MPI jobs should have a dedicated network to run those types of jobs. It was noticed in the test-run phases that MPI processes in general generate huge data, which in turn requires lots of network bandwidth. This is mainly caused by the higher speed of Multi-processing in each node in relation to the current speed available in the test cluster.

The main advantage of using Intel's MPI implementation in this work is the ability to define network or device fabrics, or in other words, defining a clusters physical connectivity. In this cluster, the fabric can be defined as a TCP network with shared-memory cores, which is synonymous to an Ethernet-based SMP cluster. When running without explicitly defining the underlying fabric of our cluster, overall performance degradation was noticeable as the cluster's overall benchmark result was merely 240 Gflops, a 40 Gflops more than the previously mentioned failed attempts with an MPI only approach but still a low number when considering the overall expected performance. This was caused by having the MPI processes started without previous knowledge of multiple-core nodes, in this scenario, each core will be treated as a single component with no communicative relation with its neighboring cores within the same node, resulting in communication rather than processing which leads to more idle time for that specific core. To solve the problem of the low overall results obtained, a new parameter to define the underlying mechanism for the running MPI library was introduced in next test-runs, and as expected, the results obtained reached a maximum of 662.6 Gflops. This was the expected result at the beginning of the test, but was not achievable with our preliminary runs since it additionally needed a definition of the underlying fabric for Intel's MPI to use in order to achieve such performance. The addition of the option lead to an execution aware of both communication types available for this cluster which are the Gigabit communication between the nodes and the shared memory communication within a node's cores. This essentially leads to the better performance achieved.

Another aspect of these tests was how the cluster was viewed or perceived physically and how that differed from the way we should look at it. When dealing with Multi-core processors, an abstract view is needed as well, and the best method was to use diagrams, such as in figure 1 and 2. These figures depict how a 6x16 topology was chosen and how processes are distributed among nodes. It can be noticed from the figures that processes are passed in a round-robin way across different cores and not nodes. In this cluster, each node has 8 processors, so it can be viewed as 8 different single-core processor nodes. This distribution of processes affects the overall performance as well. Unexpectedly, the 6x16 grid performed well as a result of having more related processors on a single node, as well as less communication is needed between the processes across the grid. In this configuration, each of the running processes can heavily utilize the shared cache and local communication bridges to accomplish some of the tasks. On the other hand, network communication happens while processing cores are being utilized for processing. Table 6 summarizes the best as well as unexpected results obtained from several test-runs. From Table 6, we can notice the drastic performance obtained by changing the way we deal with modern day computer clusters. A high increase in performance was the result of an experience we attained when dealing with this new types of clusters.

**Table 6.** General summary of trials.

| Option Types | Gflops Obtained | PxQ | Problem size N | Comments |
|---|---|---|---|---|
| OpenMPI, MPI | 207 Gflops | 8x12 | 140000 | Low results, tested with different topologies and mapping schemes. |
| Intel MPI, fabric-less | 204 Gflops | 8x12 | 140000 | Another low result, although expectations were high, using non-MPI-only network. |
| Intel MPI, fabric-less | 224.6 Gflops | 6x16 | 140000 | Good indication of the 6x16 topology which lead us to choose it in later phases. |
| Intel MPI, TCP+Shared Mem. | 662.6 Gflops | 6x16 | 140000 | The hybrid mode reaches a new peak, 60% overall efficiency. |

In general, we can summarize the main observations gathered from a modern day cluster in the following points:

1. The network significantly affects the Cluster's performance. Thus, separating the MPI network from the normal network may result in better overall performance of the cluster.
2. The cluster's compute processors and the architecture, of which the processors inherit their features from, should be studied as different processors perform differently.
3. The MPI implementation in use must be considered since not all provide the same features and perform similarly as shown in Table 6. Although all libraries can run MPI jobs, as well as the different approaches available for cluster users. An example of MPI libraries available are the OpenMPI library and the Intel MPI library implementation.
4. Both the physical and abstract aspects are important, details of how MPI applications process data must at least be known by a cluster administrator, as these details will determine how a cluster performs.
5. Multi Node cluster scalability is still debatable; scaling a cluster without upgrading network bandwidth may not achieve its goal of performance improvement as we found out in this work. Performance degradation caused by scaling up was relatively high; we assume a faster network will yield better performance in relation to scalability for these types of clusters.

## 6   Conclusion

From the obtained results, we can observe the difference between the MPI-OpenMP hybrid implementations and an MPI-only implementation. Moreover, how this may heavily affect a benchmark test when done on Multi-core clusters. The numbers obtained are the results of test-runs executed on a 12 node cluster with 96 cores. It was done for the purpose of knowing how scalable the cluster was and how good was it to perform, while that happened, many more observations were recorded that we hope will benefit the researchers and practitioners working with such clusters.